\documentclass[a4paper,11pt]{article}% %
\usepackage[width=15cm,height=23cm]{geometry}
\usepackage[usenames]{color}
\usepackage{amsfonts,amssymb}
\usepackage{theorem}
\usepackage[dvips]{graphicx}
\newcommand{\su}[2]{\stackrel{#1}{#2}}

\long\def\symbolfootnote[#1]#2{\begingroup%
\def\thefootnote{\fnsymbol{footnote}}\footnote[#1]{#2}\endgroup}

\def\G{\Gamma}

\begin{document}
%\rightline{ MIT, March 31, 2010 at 11.59 am}
%\rightline{ MIT, February 6, 2011 at 11.59 pm}
%\rightline{ferrrai\_final.tex}

%\vskip 0.4 truecm
%\Large

\begin{center}{\Large\bf
Of Higgs, Unitarity and other Questions
\footnote{Presented
by Ruggero Ferrari at the International Conference {\sl
"Gauge Fields. Yesterday, Today, Tomorrow"} in honor of A.A. Slavnov.
Moscow, January 19-24 2010.
}}
\end{center}
\par
%\normalsize
%\large
%\vskip 0.5 truecm

%\large
\rm
%\vskip 0.7 truecm
\centerline{D.~Bettinelli$^c$\footnote{e-mail:
{\tt daniele.bettinelli@physik.uni-freiburg.de}},
R.~Ferrari$^{a,b}$\footnote{e-mail: {\tt ruggero.ferrari@mi.infn.it}},
A.~Quadri$^b$\footnote{e-mail: {\tt andrea.quadri@mi.infn.it}}}

%\normalsize
\small
\medskip
\begin{center}
$^a$
Center for Theoretical Physics\\
Laboratory for Nuclear Science\\
and Department of Physics\\
Massachusetts Institute of Technology\\
Cambridge, Massachusetts 02139
and\\
$^b$
Dip. di Fisica, Universit\`a degli Studi di Milano\\
and INFN, Sez. di Milano\\
via Celoria 16, I-20133 Milano, Italy\\
$^c$
Physikalisches Institut, Albert-Ludwigs-Universit\"at Freiburg\\
Hermann-Herder-Str. 3, D-79104 Freiburg im Breisgau, Germany.\\
(MIT-CTP-4138, March, 2010 )
\end{center}

\normalsize

\vskip 1.3  truecm

%\centerline{Abstract}
\begin{quotation}

\rm
 {\Large Abstract:}
On the verge of conclusive checks on the Standard Model by the LHC,
we discuss some of the basic assumptions. The reason for this
analysis stems from a recent proposal of an Electroweak Model based
on a nonlinearly realized gauge group $SU(2)\otimes U(1)$, where, in
the perturbative approximation, there is no Higgs boson. The model
enjoys the Slavnov-Taylor identities and therefore the
perturbative unitarity. On the other hand, it is commonly believed
that the existence of the Higgs boson is entangled with the property
of unitarity, when high energy processes are considered. The
argument is based mostly on the Froissart bound and on the {\sl
Equivalence Theorem}. In this talk we briefly review some of our objections
on the validity of such arguments. Some open questions are
pointed out, in particular on the limit of zero mass for the vector
mesons and on the fate of the longitudinal polarizations.

\end{quotation}
\textheight = 233mm  \textwidth = 165mm
%%%%%%%%%%%%%%%%%%%%%%%%%%%%%%%%%%%%%%%%%%%%%%%%%%%%
%%%%%%%%%%%%%%%%%%%%%%%%%%%%%%%%%%%%%%%%%%%%%%%%%%%%
%%%%%%%%%%%%%%%%%%%%%%%%%%%%%%%%%%%%%%%%%%%%%%%%%%%%
%%%%%%%%%%%%%%%%%%%%%%%%%%%%%%%%%%%%%%%%%%%%%%%%%%%%
%%%%%%%%%%%%%%%%%%%%%%%%%%%%%%%%%%%%%%%%%%%%%%%%%%%%
\newpage
\section{Introduction}
The main assumptions for the construction of a massive Yang-Mills (YM)
\emph{local} quantum field theory are
\begin{enumerate}
\item Renormalizability
\item Unitarity
\item Spontaneous Breakdown of Symmetry.
\end{enumerate}
The mass is derived from the interaction with the Higgs field
{
\begin{eqnarray}
S_{SSB}
= S_{YM} +\frac{\Lambda^{D-4}}{g^2} \int d^Dx \frac{1}{4} Tr\Biggl\{\Bigl|\partial_\mu\Omega-i A_\mu\Omega
\Bigr|^2\Biggr\} + S_{BS} \,,
\end{eqnarray}
}
where $S_{BS}$ is the pure boson part of the action responsible for the nonzero vacuum expection 
 value of the Higgs boson field. For $SU(2)$ the  matrix $\Omega$ may be parametrized by the real fields
\begin{eqnarray}
\Omega = \phi_0 + i \tau_i \phi_i, \quad \phi_0=h+2v, \quad \langle h\rangle=0, \quad M=gv
\end{eqnarray}
%
%where $ S_{BS}$ is the pure boson part of the action responsible of the
%non-zero vacuum expectation value of the Higgs Boson field.
%%%%%%%%%%%%%%%%%%%%%%%%%%%%%%%%%%%%%%%%%%%%%%%%%%%%%%%
%%%%%%%%%%%%%%%%%%%%%%%%%%%%%%%%%%%%%%%%%%%%%%%%%%%%%%%
%%%%%%%%%%%%%%%%%%%%%%%%%%%%%%%%%%%%%%%%%%%%%%%%%%%%%%%
\par
{ In this talk we focus mainly on the issue of unitarity and on its
connection with the presence of a physical Higgs boson in the
perturbative spectrum. In part one we consider a brief statement of
the problem on general grounds, i.e. on the perturbative unitarity
and its relationship with the optical theorem. In part two we derive
 relations between the amplitudes involving the scalar
part of the vector mesons on the one hand and the Goldstone bosons on the other.
These relations are somehow related to the so-called Equivalence
Theorem in gauge theories. In part three we flash some of the work
we did on the nonlinear sigma model and on the massive Yang-Mills theory in
order to put on a subtraction procedure for these nonrenormalized
theories.

\section{Part One: Unitarity}
\par
The attention has been focused on the $W_L W_L$ elastic scattering process for different
reasons. At high energy ($s,t>>M^2_W$) some anomalous behavior is expected
for the longitudinal polarization. The idea is to entangle the
presence of the Higgs boson with the requirement of unitarity.
The calculations often make use of the so called \emph{Equivalence
Theorem} \cite{Cornwall:1974km}-\cite{Gounaris:1986cr}.
%%%%%%%%%%%%%%%%%%%%%%%%%%%%%%%%%%%%%%%%%%%%%%%%%%%%%%%
%%%%%%%%%%%%%%%%%%%%%%%%%%%%%%%%%%%%%%%%%%%%%%%%%%%%%%%
%%%%%%%%%%%%%%%%%%%%%%%%%%%%%%%%%%%%%%%%%%%%%%%%%%%%%%%
}%%

\subsection{Unitarity:}
It is worth stressing the conceptual difference between
the \emph{Optical Theorem} for the $S$-matrix
\begin{eqnarray}
S=I\!\!I-iT, \quad S^\dagger S=I\!\!I,\quad \Longrightarrow Im T_{ii} \sim \sigma_{iT}
\label{i.1}
\end{eqnarray}
and
\emph{Perturbative Unitarity}
\begin{eqnarray}
\sum_{j=0}^k S^{(j)\dagger}S^{(k-j)}=0, \quad \forall k>0,
\label{i.3}
\end{eqnarray}
where
\begin{eqnarray}
S= \sum_{k=0}^\infty S^{(k)}, \quad S^{(0)}= I\!\!I.
\label{i.2}
\end{eqnarray}
For any
finite order calculation $S_{in}= \sum_{j=0}^k S^{(j)}_{in}$
\begin{eqnarray}&&
 \sum_n \Biggl|\sum_{j=0}^k S^{(j)}_{in} \Biggr|^2
 = \sum_n \sum_{l=0}^k \sum_{j=0}^k S^{(j)*}_{in}S^{(l-j)}_{in}
 + \sum_n \sum_{l=k+1}^{2k}\sum_{j=0}^k S^{(j+l)*}_{in}S^{(l-j)}_{in}
 \nonumber\\&&
 =1 + \sum_n \sum_{l=k+1}^{2k}\sum_{j=l-k}^k S^{(j)*}_{in}S^{(l-j)}_{in}
\end{eqnarray}
There is always a violation of the Optical Theorem of
order ${\cal O}(k+1)$.
\par\noindent The Optical Theorem has a meaning only if
an operative definition of \emph{forward} scattering exists.
If long range interactions are present, then the forward amplitude
is an elusive object. Only eq. (\ref{i.3}) has a meaning.

\section{Part Two:  Equivalence Theorem}

This part is devoted to the discussion of some aspects of the massive YM theory
in the linear representation of the gauge group of local transformations
(Higgs mechanism). Most of the results are also valid for the case in which the
representation is nonlinear (St\"uckelberg mass).

\subsection{BRST  Transformations:}

The BRST differential ${\mathfrak s}$ is obtained in the usual way by promoting
the gauge parameters  to the ghost fields $c_a$ and by introducing
the antighosts $\bar c_a$ coupled in a BRST doublet to the Nakanishi-Lautrup fields $b_a$:
\begin{eqnarray}
&& {\mathfrak s} \phi_a = \frac{1}{2} \phi_0 c_a + \frac{1}{2} \epsilon_{abc} \phi_b c_c \, , ~~~~
{\mathfrak s} \phi_0 = -\frac{1}{2} \phi_a c_a
\nonumber \\
&& {\mathfrak s} A_{a\mu} = (D_\mu [A]c)_a \quad , {\mathfrak s} \bar c_a = b_a, \quad {\mathfrak s} b_a = 0 \, .
\label{brst.1}
\end{eqnarray}
In the above equation $D_\mu[A]$ denotes the covariant derivative w.r.t. $A_{a\mu}$:
\begin{eqnarray}
(D_\mu[A])_{ac} = \delta_{ac} \partial_\mu + \epsilon_{abc} A_{b\mu} \, .
\label{brst.1.1}
\end{eqnarray}
The BRST transformation of $c_a$ then follows by nilpotency,
\begin{eqnarray}
{\mathfrak s} c_a = - \frac{1}{2} \epsilon_{abc} c_b c_c \, .
\label{brst.2}
\end{eqnarray}

%%%%%%%%%%%%%%%%%%%%%%%%%%%%%%%%%%%%%%%%%%%%%%%%%
%%%%%%%%%%%%%%%%%%%%%%%%%%%%%%%%%%%%%%%%%%%%%%%%%
%%%%%%%%%%%%%%%%%%%%%%%%%%%%%%%%%%%%%%%%%%%%%%%%%

The tree-level vertex functional  is
\begin{eqnarray}
\G^{(0)} & = & S_{SSB} +\frac{\Lambda^{(D-4)}}{g^2}~ {\mathfrak s}~ \int d^Dx \,
( \bar c_a \partial A_a )
\nonumber \\&&
+ \frac{\Lambda^{(D-4)}}{g^2}~\int d^Dx \, (A_{a\mu}^*  {\mathfrak s} A_a^\mu
+ \phi_a^* {\mathfrak s} \phi_a + \phi_0^* {\mathfrak s} \phi_0
+ c_a^* {\mathfrak s} c_a ) \nonumber \\
& = & S_{YM} + \frac{\Lambda^{(D-4)}}{g^2}\int d^Dx \, \Big ( b_a \partial A_a - \bar c_a \partial_\mu (D^\mu[A] c)_a \Big ) \nonumber \\
&   & +\frac{\Lambda^{(D-4)}}{g^2}~\int d^Dx \, (A_{a\mu}^*  {\mathfrak s} A_a^\mu + \phi_0^* {\mathfrak s} \phi_0
+ \phi_a^* {\mathfrak s} \phi_a + c_a^* {\mathfrak s} c_a ) \, .
\label{brst.3}
\end{eqnarray}
In $\G^{(0)}$ we have also included the antifields $A_{a\mu}^*,
\phi_0^*, \phi_a^*$ and $c_a^*$ coupled to the nonlinear BRST
variations of the quantized fields.

%%%%%%%%%%%%%%%%%%%%%%%%%%%%%%%%%%%%%%%%%%%%%%%%%
%%%%%%%%%%%%%%%%%%%%%%%%%%%%%%%%%%%%%%%%%%%%%%%%%
%
%%%%%%%%%%%%%%%%%%%%%%%%%%%%%%%%%%%%%%%%%%%%%%%%
\subsection{Slavnov-Taylor Identity (STI):}
To simplify notations, we perform the substitution $ b_a \to
\frac{g^2}{\Lambda^{(D-4)}}b_a $.
%
%\par
The STI are
%\par\noindent
for the 1-PI functional (ZJ renormalization of composite operators) is
\begin{eqnarray}
\!\!\!\!\!\!\!\!
\int d^Dx \, \Big (
 \G_{ A^*_{a\mu}}  \G_{ A_a^\mu}
+
 \G_{ \phi_a^*}  \G_{ \phi_a}
+
 \G_{ \phi_0^*}  \G_{ \phi_0}
+
 \G_{ c_a^*} \G_{ c_a}
 +b_a  \G_{ \bar c_a}
%+ \frac{\Lambda^{(D-4)}}{g^2}b_a  \G_{ \bar c_a}
\Big ) = 0 \, ,
\label{brst.13}
\end{eqnarray}
where we use the notation
\begin{eqnarray}
\Gamma_X\equiv  \frac{\delta \Gamma}{\delta X},
%\label{YM.8}
\end{eqnarray}
%
%%%%%%%%%%%%%%%%%%%%%%%%%%%%%%%%%%%%%%%%%%%%%%%%
%%%%%%%%%%%%%%%%%%%%%%%%%%%%%%%%%%%%%%%%%%%%%%%%%
%%%%%%%%%%%%%%%%%%%%%%%%%%%%%%%%%%%%%%%%%%%%%%%%%
while  for the generating functional of the connected amplitudes
one has
\begin{eqnarray}
\!\!\!\!\!
\!\!\!\!\!
\int d^Dx \, \Big (
- W_{ A^*_{a\mu}}J_{a\mu}
-
 W_{ \phi_a^*} K_a
-  W_{ \phi_0^*}K_0
+
 W_{ c_a^*}\bar\eta_a
- W_{  b_a} \eta_a
\Big ) = 0 \,
\label{YM.10}
\end{eqnarray}
We use the notations
\begin{eqnarray}
W_{ A^*_{a\mu}\dots}\equiv  \frac{\delta^n W}{\delta A^*_{a\mu} \dots}
=  i^{n-1}\langle 0|T((D^\mu [A]c)_a\dots)|0\rangle_C
%\label{YM.9}
\end{eqnarray}
for  composite fields,
 while for elementary fields
\begin{eqnarray}
W_{ \underbrace{b_a\dots}_n}\equiv  i^{n-1}\langle 0|T(b_a\dots)|0\rangle_C.
%\label{YM.10.1}
\end{eqnarray}
The external field sources are 
\begin{eqnarray}
&&
\int d^Dx \, \Big (J_{a\mu}A^\mu_a+ K_a\phi_a+ K_0\phi_0+ \bar c_a\eta_a
+\bar\eta_a c_a + J_{b_a}b_a
\Big )  \, .
\end{eqnarray}
%
%%%%%%%%%%%%%%%%%%%%%%%%%%%%%%%%%%%%%%%%%
%%%%%%%%%%%%%%%%%%%%%%%%%%%%%%%%%%%%%%%%%
 %%%%%%%%%%%%%%%%%%%%%%%%%%%%%%%%%%%%%%%%%%%%%%%%
%%%%%%%%%%%%%%%%%%%%%%%%%%%%%%%%%%%%%%%%%%%%%%%%%
%%%%%%%%%%%%%%%%%%%%%%%%%%%%%%%%%%%%%%%%%%%%%%%%%

\subsection*{Landau Gauge Equation}
The equation associated to the gauge fixing
gives
\begin{eqnarray}
\Gamma_{b_a}
=
 \partial_\nu A_a^\nu
\label{YM.11}
\end{eqnarray}
\begin{eqnarray}
-J_{b_a}
=
 \partial_\mu W_{ A_a^\mu}.
\label{YM.12}
\end{eqnarray}
The antighost equations can be derived from eqs. (\ref{brst.13}), (\ref{YM.10}),
(\ref{YM.11}) and (\ref{YM.12}):
\begin{eqnarray}
\Gamma_{\bar c_a}
= \partial_\nu \Gamma_{A^*{a\nu}}
\label{YM.13}
\end{eqnarray}
\begin{eqnarray}
\eta = \partial^\nu W_{ A_a^{*\nu}}.
\label{YM.14}
\end{eqnarray}
From eqs. (\ref{YM.10}) and (\ref{YM.14}) one gets
\begin{eqnarray}&&
W_{\phi^* \bar c} = W_{\phi  b}
\nonumber\\&&
W_{A_{\mu}^* \bar c} = W_{A^\mu  b}=-i \frac{p^\mu}{p^2}.
\label{YM.15}
\end{eqnarray}

\subsection*{Some Basic Results}
By a straightforward use of the above equations and of
\begin{eqnarray}
\Gamma W = - I\!\!I,
%\label{basic.7}
\end{eqnarray}
one gets
\begin{eqnarray}
 W_{\phi b}
=i\frac{p^\nu\Gamma_{\phi A^\nu}}{\Gamma_{\phi\phi}}
\frac{1}{p^2}
%\label{tpf.9}
\end{eqnarray}
\begin{eqnarray}
%\fbox
{
(p^\nu\Gamma_{A^\nu\phi})^2 +p^2\Gamma_L\Gamma_{\phi\phi}=0
}
%\label{tpf.9.2}
\end{eqnarray}
\begin{eqnarray}
  W_{A^\mu \phi}=0,\quad
W_L
=0,\quad
W_{\phi \phi}
 =-
\frac{1}{ \Gamma_{\phi \phi}}.
\label{tpf.22}
\end{eqnarray}

\subsection*{Free Fields}
The 2-point 1-PI functions are given by
\begin{eqnarray}&&
\Gamma_{ A^\nu\phi}=iMp_\nu, \,\,
\Gamma_{bb}= 0, \,\,
\Gamma_{ A^\nu b}=ip_\nu, \,\,
\nonumber\\&&
\Gamma_{\phi\phi}=p^2, \,\,
\Gamma_{\phi b }=0, \,\,
\Gamma_L=M^2.
%\label{tpf.25}
\end{eqnarray}
Then
\begin{eqnarray}&&
 W_{A^\mu\phi }=0
\label{tpf.26}
\end{eqnarray}
and
\begin{eqnarray}&&
 W_{A^\mu b}=-i\frac{p^\mu}{
p^2
},\qquad
 W_{\phi b}=\frac{M}{
p^2
}
\nonumber\\&&
 W_L=0,\qquad
W_{\phi \phi}
 =- \frac{1}{
p^2
}.
\label{tpf.27}
\end{eqnarray}
\textbf{Theorem} For $m\ge 1$
\begin{eqnarray}&&
W_{b_{x_1}\cdots b_{x_m}}=0 \,\, ,
\nonumber\\&&
W_{b_{x_1}\cdots b_{x_m} \phi^*_{z_1}\bar c_{y_1}\cdots \phi^*_{z_k}\bar c_{y_k} }=0\,\, ,
\nonumber\\&&
W_{b_xb_{x_1}\cdots b_{x_m} \phi_{w_1}\cdots\phi_{w_n}}
=
\sum_{i=1}^n W_{\phi^*_{w_i}\bar c_x b_1\cdots b_m \phi_{w_1}\cdots\su{\vee}{\phi}_{w_i}\cdots\phi_{w_n}}\,\,  ,
\nonumber\\&&
 \sum_{j=1}^k ~(-1)^j~W_{b_{y_j} b_{x_1}\cdots b_{x_m}\phi^*_{z_1}\cdots\phi^*_{z_{k-1}}
\bar c_{y_1}\cdots\su{\vee}{\bar c}_{y_j}\cdots \bar  c_{y_k}
\phi_{w_1}\cdots\phi_{w_n}}
\nonumber\\&&
+\sum_{i=1}^n
W_{ b_{x_1}\cdots b_{x_m}
\phi^*_{z_1} \cdots \phi^*_{z_{k-1}}\phi^*_{w_i}\bar c_{y_1}\cdots \bar  c_{y_k}
\phi_{w_1}\cdots\su{\vee}{\phi}_{w_i}\cdots\phi_{w_n}}
=0
\label{basic.19}
\end{eqnarray}
where $\su{\vee}{}$ marks omitted symbols. Proof: just use  the STI.
\par\noindent
Eq. (\ref{basic.19}) is easily generalized  to the case where any number of external physical
legs are added (via the reduction formulae formalism).

\subsection{$b$-insertions}
The quantity
\begin{eqnarray}
R\equiv i\frac{p^\nu\Gamma_{\phi A^\nu}}{M\Gamma_{\phi\phi}}\Big|_{p^2=0}
=\frac{p^2}{M} W_{b\phi}\Big|_{p^2=0}
\label{basic.11}
\end{eqnarray}
will appear all over again (at the tree level $R=1$). The pole contribution gives
\begin{eqnarray}&&
\lim_{p^2=0}p^2 W_{b(p) \cdots}=\biggl(
i\frac{p^\nu\Gamma_{\phi A^\nu}}{\Gamma_{\phi\phi}}W_{\widehat{\phi(p)} \cdots}
+ip^\mu W_{\widehat{A^\mu(p)} \cdots}\biggr)\Big |_{p^2=0}
\nonumber\\&&
=\biggl( - M R \Gamma_{\phi\phi}W_{\phi(p) \cdots}
+ip^\mu W_{\widehat{A^\mu(p)} \cdots}\biggr)\Big |_{p^2=0}
.
\label{basic.9}
\end{eqnarray}
%
%%%%%%%%%%%%%%%%%%%%%%%%%%%%%%%%%%%%%%%%%
%%%%%%%%%%%%%%%%%%%%%%%%%%%%%%%%%%%%%%%%%
Then one $b$-insertion on a physical amplitude yields
\begin{eqnarray}&&
\lim_{p^2=0}p^2 W_{b(p) ***}=\biggl(
i\frac{p^\nu\Gamma_{\phi A^\nu}}{\Gamma_{\phi\phi}}W_{\widehat{\phi(p)} ***}
+ip^\mu W_{\widehat{A^\mu(p)} ***}\biggr)\Big |_{p^2=0}
\nonumber\\&&
=\biggl( - M R \Gamma_{\phi\phi}W_{\phi(p) ***}
+ip^\mu W_{\widehat{A^\mu(p)} ***}\biggr)\Big |_{p^2=0}
=0,
\label{basic.9.1}
\end{eqnarray}
where the $\scriptstyle{ ***}$ indicates all the other \emph{physical} states obtained
via reduction formulas.

The~ $\widehat{}$~ indicates that the external line
(for instance, attached to an $A^\mu$) has been removed. According to this
notation
\begin{eqnarray}
W_{A(p)BC\dots}= \sum_X W_{A(p)X} W_{\widehat{X(p})BC\dots}.
\label{basic.10}
\end{eqnarray}

\subsection*{The Longitudinal Polarization}
The relation with the longitudinal polarization
\begin{eqnarray}
\epsilon_L = \frac{E}{M |\vec p|}\Big(\frac{\vec p\,^2}{E}, \vec p\Big),
\qquad E=\sqrt{M^2+\vec p\,^2}
\label{basic.11.1}
\end{eqnarray}
can be obtained by considering
{
\begin{eqnarray}
\epsilon_L = \frac{E}{M |\vec p|}\Big(|\vec p|, \vec p\Big)
-  \frac{M}
{E+|\vec p|}(1, \vec 0 ).
\label{basic.11.2}
\end{eqnarray}
}%%
It is usually  assumed that
\begin{eqnarray}
\epsilon_L = \frac{1}{M}\Big(|\vec p|, \vec p\Big)
+ {\cal O}(\frac{M}{E})
\label{basic.11.3}
\end{eqnarray}
gives the correct order of magnitude in the amplitudes
\begin{eqnarray}
\epsilon_L^\mu W_{\widehat{A^\mu(p)} ***}\biggr|_{p^2=M^2} = \frac{1}{M}p^\mu W_{\widehat{A^\mu(p)} ***}\biggr|_{p^2=0}
+ {\cal O}(\frac{M}{E})\,.
\label{basic.11.4}
\end{eqnarray}
Then eq. (\ref{basic.9}) reads
\begin{eqnarray}
\epsilon_L^\mu W_{\widehat{A^\mu(p)} ***}\biggr|_{p^2=M^2} =
i R W_{\widehat{\phi(p)} ***}\biggr|_{p^2=0}
+ {\cal O}(\frac{M}{E})\,,
\label{basic.11.4.1}
\end{eqnarray}
which is the statement of Lee, Quigg, Thacker (1977)\cite{Lee:1977eg}, Weldon (84)\cite{Weldon:1984wt},
Chanowitz, Gaillard (1985)\cite{Chanowitz:1985hj},
Gounaris, K\"ogerler, Neufeld (1986)\cite{Gounaris:1986cr}.
\par
{
Unfortunately, it will appear that the evaluation of the order of magnitude
given in eq. (\ref{basic.11.3}) cannot always be transfered to the amplitudes
as indicated by eq.(\ref{basic.11.4}).
In particular, there is a clear evidence that the limit $M=0$
does not commute with the on-shell limit (reduction formula) as shown by the
example with two $b-$insertions below.
}%%

\subsection*{Two $b$-insertions}
This is a very clear example of the singular behavior of the limit
$M=0$. The situation is somewhat different if we use eq.
(\ref{basic.9}) or (\ref{basic.11.4}). We use first eq.
(\ref{basic.9}) and subsequently we discuss the approach by
exploiting eq. (\ref{basic.11.4}). {Note that the insertion of a second
$b_a$ line is much simpler in the Landau gauge where $W_{A\phi} =0$
remains valid beyond the tree approximation. In generic 't Hooft
gauge there is a non-trivial mixing in the $\phi- \partial_\mu
A^\mu$ space, which causes some important technical complexities.}
\par\noindent
One has
\begin{eqnarray}&&
\lim_{p_1^2,p_2^2=0}i^2 p_1^\mu p_2^\nu W_{\widehat{A^\mu(p_1)}\widehat{A^\nu(p_2)} ***}
= \lim_{p_1^2,p_2^2=0}p_1^2
p_2^2\biggl( W_{b_1b_2 ***}
\nonumber\\&&
+\frac{ip_1^\nu \Gamma_{\phi A^\nu}}{p_1^2}  W_{\phi_1b_2 ***}
+\frac{ip_2^\mu \Gamma_{\phi A^\mu}}{p_2^2}W_{b_1\phi_2 ***}
+\frac{(ip_1^\nu \Gamma_{\phi A^\nu} )}{p_1^2}\frac{(ip_2^\mu \Gamma_{\phi A^\mu})}{p_2^2} W_{\phi_1\phi_2 ***}\biggr)
\nonumber\\&&
\label{basic.12}
\end{eqnarray}
The first term is zero as in eq. (\ref{basic.19}).
%
 %%%%%%%%%%%%%%%%%%%%%%%%%%%%%%%%%%%%%%%%%%%%%%%%
%%%%%%%%%%%%%%%%%%%%%%%%%%%%%%%%%%%%%%%%%%%%%%%%%
%%%%%%%%%%%%%%%%%%%%%%%%%%%%%%%%%%%%%%%%%%%%%%%%%
 The mixed terms
can be obtained by performing the functional derivatives of the STI
in eq. (\ref{YM.10}) with respect to $\eta$ and $K$,
\begin{eqnarray}
W_{b_1\phi_2 ***} = W_{\phi^*_2  \bar c_1 ***}.
\label{basic.13}
\end{eqnarray}
Thus we get (with the use of $W_{\phi^*  \bar c_1}=W_{b\phi}$)
\begin{eqnarray}&&
\lim_{p_1^2,p_2^2=0}i^2 p_1^\mu p_2^\nu
 W_{\widehat{A^\mu(p_1)}\widehat{A^\nu(p_2)} ***}
=  M^2 R^2\lim_{p_1^2,p_2^2=0} \biggl(
\frac{\Gamma_{\phi_1\phi_1}} {p_1^2} p_2^2
W_{ \bar c_2 c_2}
W_{\widehat{ \,\,\bar c_1}\widehat{\,\,   c_2} ***}
\nonumber\\&&
+\frac{\Gamma_{\phi_2\phi_2}} {p_2^2} p_1^2
W_{ \bar c_1 c_1}
W_{\widehat{ \,\, \bar c_2}\widehat{\,\,   c_1} ***}
+ W_{\widehat{\phi_1}\widehat{\phi_2} ***}\biggr)
\nonumber\\&&
= M^2 R^2\lim_{p_1^2,p_2^2=0} \biggl(
\bar R
W_{\widehat{ \,\,\bar c_1}\widehat{\,\,   c_2} ***}
%\nonumber\\&&
+\bar R
W_{\widehat{ \,\, \bar c_2}\widehat{\,\,   c_1} ***}
+ W_{\widehat{\phi_1}\widehat{\phi_2} ***}\biggr),
\label{basic.14}
\end{eqnarray}
where
\begin{eqnarray}
\bar R=
\lim_{p_1^2,p_2^2=0}
\frac{\Gamma_{\phi_2\phi_2}} {p_2^2} p_1^2
W_{ \bar c_1 c_1}.
\label{basic.15}
\end{eqnarray}
%
%
%%%%%%%%%%%%%%%%%%%%%%%%%%%%%%%%%%%%%%%%%%%%%%%%
%%%%%%%%%%%%%%%%%%%%%%%%%%%%%%%%%%%%%%%%%%%%%%%%%
%%%%%%%%%%%%%%%%%%%%%%%%%%%%%%%%%%%%%%%%%%%%%%%%%
%
On the other hand, if we consider multi $b$-field insertions by
using eq. (\ref{basic.9}), where the scalar mode is replaced by the
longitudinal mode according to eq. (\ref{basic.11.4}),
\begin{eqnarray}
RW_{\widehat{\phi(p)} ***} =
\lim_{p^2=0}\frac{p^2}{M} W_{b(p) ***}- i\epsilon_L^\mu
 W_{\widehat{A^\mu(p)} ***}\Big |_{p^2=M^2}+ {\cal O}(\frac{M}{E}),
\label{basic.16}
\end{eqnarray}
we get
\begin{eqnarray}&&
R^2W_{\widehat{\phi(p_1)} \widehat{\phi(p_2)}***} =
\lim_{p_1^2,p_2^2=0}\frac{p_1^2 p_2^2}{M^2} W_{b(p_1)b(p_2) ***}
+i\lim_{p_1^2=0}\frac{p_1^2 }{M}\epsilon_L^{\mu_2}W_{b(p_1)\widehat{A^{\mu_2}(p_2)} ***}\Big |_{p_2^2=M^2}
\nonumber\\&&
+i\lim_{p_2^2=0}\frac{p_2^2 }{M}\epsilon_L^{\mu_1}W_{\widehat{A^{\mu_1}(p_1)}b(p_2) ***}\Big |_{p_1^2=M^2}
+ i^2\epsilon_L^{\mu_1}
\epsilon_L^{\mu_2}
 W_{\widehat{A^{\mu_1}(p_1)} \widehat{A^{\mu_2}(p_2)}***}\Big |_{
p_1^2,p_2^2=M^2}
+ {\cal O}(\frac{M}{E})
\nonumber\\&&
 =
 -\epsilon_L^{\mu_1}
\epsilon_L^{\mu_2}
 W_{\widehat{A^{\mu_1}(p_1)} \widehat{A^{\mu_2}(p_2)}***}\Big |_{
p_1^2,p_2^2=M^2}+ {\cal O}(\frac{M}{E}),
\label{basic.17}
\end{eqnarray}
where the mixed terms and the double $b$-insertion are zero as required by eq. (\ref{basic.19}).
%
 %%%%%%%%%%%%%%%%%%%%%%%%%%%%%%%%%%%%%%%%%%%%%%%%
%%%%%%%%%%%%%%%%%%%%%%%%%%%%%%%%%%%%%%%%%%%%%%%%%
%%%%%%%%%%%%%%%%%%%%%%%%%%%%%%%%%%%%%%%%%%%%%%%%%
By replacing the scalar mode (unphysical) with the longitudinal polarization state,
the value of the $b$-insertions changes in a substantial way.
\par
We can conclude that the use of the substitution in eq.
(\ref{basic.9}) brings in a  contradiction between the results in
eqs. (\ref{basic.14}) and (\ref{basic.17}). This fact has been
pointed out in Ref. \cite{Chanowitz:1985hj}.

\subsection*{Three $b$-insertions}
%%%%%%%%%%%%%%%%%%%%%%%%%%%%%%%%%%%%%%%%%
We consider three $b$-insertions, which can be relevant in processes
 like $V+V\to l^+ + l^- + V$. We use once again the eq. (\ref{basic.9})
as in eq. (\ref{basic.12}):
\begin{eqnarray}&&
\lim_{p_1^2,p_2^2,p_3^2=0}\frac{i^3}{M^3} p_1^\mu p_2^\nu p_3^\rho
W_{\widehat{A^\mu(p_1)}\widehat{A^\nu(p_2)}\widehat{A^\rho(p_3)} ***}
\label{basic.18.0}
\end{eqnarray}

\begin{eqnarray}&&
= \lim_{p_1^2,p_2^2,p_3^2=0}p_1^2
p_2^2p_3^2\Biggl( \frac{1}{M^3}W_{b_1b_2b_3 ***}
+\frac{R \Gamma_{\phi_1\phi_1}}{M^2p_1^2}  W_{\phi_1b_2b_3 ***}
\nonumber\\&&
+\frac{R \Gamma_{\phi_2\phi_2}}{M^2p_2^2}W_{b_1\phi_2 b_3***}
+\frac{R \Gamma_{\phi_3\phi_3}}{M^2p_3^2}W_{b_1b_2\phi_3***}
\nonumber\\&&
+\frac{1}{M}\frac{R \Gamma_{\phi_1\phi_1}}{p_1^2}\frac{R \Gamma_{\phi_2\phi_2}}{p_2^2}  W_{\phi_1\phi_2b_3 ***}
+\frac{1}{M}\frac{R \Gamma_{\phi_2\phi_2}}{p_2^2}\frac{R\Gamma_{\phi_3\phi_3}}{p_3^2}W_{b_1\phi_2 \phi_3***}
\nonumber\\&&
+\frac{1}{M}\frac{R \Gamma_{\phi_3\phi_3}}{p_3^2}\frac{R \Gamma_{\phi_1\phi_1}}{p_1^2}W_{\phi_1b_2\phi_3***}
+ \frac{R \Gamma_{\phi_1\phi_1}}{p_1^2}\frac{R \Gamma_{\phi_2\phi_2}}{p_2^2}
\frac{R \Gamma_{\phi_3\phi_3}}{p_3^2}
W_{\phi_1\phi_2\phi_3 ***}\Biggr)
\nonumber\\&&
\label{basic.18}
\end{eqnarray}
The mixed terms in eq. (\ref{basic.18}) are evaluated by using eq. (\ref{basic.19}):
\begin{eqnarray}&&
\lim_{p_1^2,p_2^2,p_3^2=0}\frac{i^3}{M^3} p_1^\mu p_2^\nu p_3^\rho
W_{\widehat{A^\mu(p_1)}\widehat{A^\nu(p_2)}\widehat{A^\rho(p_3)} ***}
= \lim_{p_1^2,p_2^2,p_3^2=0}p_1^2
p_2^2p_3^2
\nonumber\\&&
\biggl(
\frac{1}{M}\frac{R \Gamma_{\phi\phi}}{p_1^2}\frac{R \Gamma_{\phi\phi}}{p_2^2}  W_{\phi_1\phi_2b_3 ***}
+\frac{1}{M}\frac{R \Gamma_{\phi\phi}}{p_2^2}\frac{R \Gamma_{\phi\phi}}{p_3^2}W_{b_1\phi_2 \phi_3***}
\nonumber\\&&
+\frac{1}{M}\frac{R \Gamma_{\phi\phi}}{p_3^2}\frac{R \Gamma_{\phi\phi}}{p_1^2}W_{\phi_1b_2\phi_3***}
+ \frac{R \Gamma_{\phi\phi}}{p_1^2}\frac{R \Gamma_{\phi\phi}}{p_2^2}
\frac{R \Gamma_{\phi\phi}}{p_3^2}
W_{\phi_1\phi_2\phi_3 ***}\biggr)
\nonumber\\&&
= -R^3\biggl(
\bar R W_{\widehat{\phi_1}\widehat{\,\,\bar c_2}\widehat{\,\,c_3} ***}
+\bar R W_{\widehat{\,\,\bar c_1}\widehat{\phi_2}\widehat{\,\,c_3} ***}
+\bar R W_{\widehat{\,\,\bar c_2}\widehat{\phi_3}\widehat{\,\,c_1} ***}
\nonumber\\&&
+\bar R W_{\widehat{\phi_2}\widehat{\,\,\bar c_3}\widehat{\,\,c_1} ***}
+\bar R W_{\widehat{\,\,\bar c_3} \widehat{\phi_1}\widehat{\,\,c_2}***}
+\bar R W_{\widehat{\phi_3} \widehat{\,\,\bar c_1}\widehat{\,\,c_2}***}
+
W_{\widehat{\phi_1}\widehat{\phi_2}\widehat{\phi_3} ***}\biggr)
\nonumber\\&&
\label{basic.20}
\end{eqnarray}

\subsection*{Four $b$-insertions}
%%%%%%%%%%%%%%%%%%%%%%%%%%%%%%%%%%%%%%%%%
There is a surprising cancellation in the case of four $b$-insertions.
\begin{eqnarray}&&
\lim_{p_1^2\dots p_4^2=0} p_1^\mu p_2^\nu p_3^\sigma p_4^\rho
W_{\widehat{A_\mu}\widehat{A_\nu}\widehat{A_\sigma}\widehat{A_\rho}}
=\lim_{p_1^2\dots p_4^2=0}  p_1^2 p_2^2 p_3^2p_4^2
\nonumber\\&&
\Biggl(  W_{b_1b_2b_3b_4}
+ M R\sum_j  \frac{\Gamma_{\phi_j\phi_j}}{p_j^2}
 W_{b_{j+1}b_{j+2}b_{j+3}{\phi_j}}
\nonumber\\&&
+ \frac{1}{2}M^2 R^2\sum_{i\not=j}  \frac{\Gamma_{\phi_k\phi_k}}{p_k^2}\frac{\Gamma_{\phi_l\phi_l}}{p_l^2}
W_{b_ib_j\,\,{\phi_k}\,\,{\phi_l}}
\nonumber\\&&
+M^3 R^3\sum_j
\frac{\Gamma_{\phi_{j+1}\phi_{j+1}}}{p_{j+1}^2}
\frac{\Gamma_{\phi_{j+2}\phi_{j+2}}}{p_{j+2}^2}
\frac{\Gamma_{\phi_{j+3}\phi_{j+3}}}{p_{j+3}^2}
W_{b_j\,\,{\phi_{j+1}}\,\,{\phi_{j+2}}\,\,{\phi_{j+3}}}
\Biggr)
\nonumber\\&&
+M^4R^4\lim_{p_1^2\dots p_4^2=0}
 W_{\widehat{\phi_{1}}\,\,\widehat{\phi_{2}}\,\,\widehat{\phi_{3}}\,\,
\widehat{\phi_{4}}}
\label{four.17}
\end{eqnarray}
%
%
 %%%%%%%%%%%%%%%%%%%%%%%%%%%%%%%%%%%%%%%%%%%%%%%%
%%%%%%%%%%%%%%%%%%%%%%%%%%%%%%%%%%%%%%%%%%%%%%%%%
%%%%%%%%%%%%%%%%%%%%%%%%%%%%%%%%%%%%%%%%%%%%%%%%%
%
Now according to the eq. (\ref{basic.19}) we have
\begin{eqnarray}
W_{b_1b_2b_3b_4}=0
\label{four.18}
\end{eqnarray}
and
\begin{eqnarray}&&
W_{b_c b_b b_a\phi_1}
=
W_{b_c b_b\phi^*_1\bar c_a}.
\label{four.20}
\end{eqnarray}
Further use of eq. (\ref{basic.19}) tells that
\begin{eqnarray}&&
W_{b_c b_b b_a\phi_1}
=
W_{b_c b_b\phi^*_1\bar c_a}=0.
\label{four.20.1}
\end{eqnarray}
We deal with the term with one $b$-insertion before considering the
most difficult term. We have again from eq. (\ref{basic.19})
\begin{eqnarray}
W_{b_j\,\,{\phi_{j+1}}\,\,{\phi_{j+2}}
\,\,{\phi_{j+3}}}
=\sum_{k=1,2,3}W_{{\phi^*_{j+k}}\,\,\bar c_j\,\,{\phi_{j+k+1}}
\,\,{\phi_{j+k+2}}}
\label{four.19.1}
\end{eqnarray}
%
 %%%%%%%%%%%%%%%%%%%%%%%%%%%%%%%%%%%%%%%%%%%%%%%%
%%%%%%%%%%%%%%%%%%%%%%%%%%%%%%%%%%%%%%%%%%%%%%%%%
%%%%%%%%%%%%%%%%%%%%%%%%%%%%%%%%%%%%%%%%%%%%%%%%%
%
Thus the relevant term in eq. (\ref{four.17}) becomes
\begin{eqnarray}&&
\lim_{p_1^2\dots p_4^2=0}p_1^2 p_2^2 p_3^2p_4^2 M^3 R^3\sum_j
\frac{\Gamma_{\phi_{j+1}\phi_{j+1}}}{p_{j+1}^2}
\frac{\Gamma_{\phi_{j+2}\phi_{j+2}}}{p_{j+2}^2}
\frac{\Gamma_{\phi_{j+3}\phi_{j+3}}}{p_{j+3}^2}
W_{b_j\,\,{\phi_{j+1}}\,\,{\phi_{j+2}}\,\,{\phi_{j+3}}}
\nonumber\\&&
=\lim_{p_1^2\dots p_4^2=0}p_1^2 p_2^2 p_3^2p_4^2 M^3 R^3\sum_j
\frac{\Gamma_{\phi_{j+1}\phi_{j+1}}}{p_{j+1}^2}
\frac{\Gamma_{\phi_{j+2}\phi_{j+2}}}{p_{j+2}^2}
\frac{\Gamma_{\phi_{j+3}\phi_{j+3}}}{p_{j+3}^2}
\nonumber\\&&
\sum_{k=1,2,3}W_{{\phi^*_{j+k}}\,\,\bar c_j\,\,{\phi_{j+k+1}}
\,\,{\phi_{j+k+2}}}
\nonumber\\&&
=\lim_{p_1^2\dots p_4^2=0}p_1^2 p_2^2 p_3^2p_4^2 M^3 R^3\sum_j
\frac{\Gamma_{\phi_{j+1}\phi_{j+1}}}{p_{j+1}^2}
\frac{\Gamma_{\phi_{j+2}\phi_{j+2}}}{p_{j+2}^2}
\frac{\Gamma_{\phi_{j+3}\phi_{j+3}}}{p_{j+3}^2}
\nonumber\\&&
\!\!\!\!\!\!\!\!\!\!\!\!\!
\sum_{k=1,2,3}W_{{\phi^*}\,\,\bar c}(p_{j+k})W_{{\bar c c}}(p_{j})W_{\phi \phi}({p_{j+k+1}})W_{\phi \phi}({p_{j+k+2}})
W_{{\widehat{\,\,\bar c_{j+k}}}\widehat{\,\, c_j}\widehat{\,\,{\phi_{j+k+1}}}\widehat{
\,\,{\phi_{j+k+2}}}}
\nonumber\\&&
=M^4 R^4\bar R\sum_j
\sum_{k=1,2,3}
W_{{\widehat{\,\,\bar c_{j+k}}}\widehat{\,\, c_j}\widehat{\,\,{\phi_{j+k+1}}}\widehat{
\,\,{\phi_{j+k+2}}}}
\label{four.19.1.1}
\end{eqnarray}
%
 %%%%%%%%%%%%%%%%%%%%%%%%%%%%%%%%%%%%%%%%%%%%%%%%
%%%%%%%%%%%%%%%%%%%%%%%%%%%%%%%%%%%%%%%%%%%%%%%%%
%%%%%%%%%%%%%%%%%%%%%%%%%%%%%%%%%%%%%%%%%%%%%%%%%
%
\par
Now we consider the most critical term in eq. (\ref{four.17}). By
using eq. (\ref{basic.19}) we get
\begin{eqnarray}
W_{b_ib_j\,\,{\phi_k}\,\,{\phi_l}}
=
W_{{\phi^*_{k}}\,\,\bar c_ib_j\,\,{\phi_{l}}}
+
W_{{\phi^*_{l}}\,\,\bar c_ib_j\,\,{\phi_{k}}}
\label{four.19.5}
\end{eqnarray}
Unfortunately, one cannot remove further the $b$-insertion by using
eq. (\ref{basic.19}). In the on-shell limit we re-express $b$ in
terms of $\phi$ and $\partial^\mu A_\mu$ as in the single pole
contribution of eq. (\ref{basic.9}). On-shell we have from eq.
(\ref{four.19.5})
\begin{eqnarray}&&
\lim_{p_j^2=0}
p_j^2
W_{b_ib_j\,\,{\phi_k}\,\,{\phi_l}}
%\nonumber\\&&
=\lim_{p_j^2=0}\Biggl [MR
\biggl(
W_{{\phi^*_{k}}\,\,\bar c_i\,\,\widehat{\phi_j}\,\,{\phi_{l}}}
+
W_{{\phi^*_{l}}\,\,\bar c_i\,\,\widehat{\phi_j}\,\,{\phi_{k}}}
\biggr )
\nonumber\\&&
+ig^2p_j^\mu \biggl(
W_{{\phi^*_{k}}\,\,\bar c_i\,\,\widehat{A^\mu_j}\,\,{\phi_{l}}}
+
W_{{\phi^*_{l}}\,\,\bar c_i\,\,\widehat{A^\mu _j}\,\,{\phi_{k}}}
\biggr)
\Biggr]
\label{four.19.6}
\end{eqnarray}
%
 %%%%%%%%%%%%%%%%%%%%%%%%%%%%%%%%%%%%%%%%%%%%%%%%
%%%%%%%%%%%%%%%%%%%%%%%%%%%%%%%%%%%%%%%%%%%%%%%%%
%%%%%%%%%%%%%%%%%%%%%%%%%%%%%%%%%%%%%%%%%%%%%%%%%
%
Thus the relevant terms in eq. (\ref{four.17}) yield
\begin{eqnarray}&&
\lim_{p_1^2\dots p_4^2=0}  p_1^2 p_2^2 p_3^2p_4^2
\frac{1}{2}M^2 R^2\sum_{i\not=j}  \frac{\Gamma_{\phi_k\phi_k}}{p_k^2}\frac{\Gamma_{\phi_l\phi_l}}{p_l^2}
W_{b_ib_j\,\,{\phi_k}\,\,{\phi_l}}
\nonumber\\&&
=\lim_{p_1^2\dots p_4^2=0}
\frac{1}{2}M^2 R^2\sum_{i\not=j}p_i^2 \Biggl(\Gamma_{\phi_k\phi_k}\Gamma_{\phi_l\phi_l}
\Biggl [
\nonumber\\&&
MR
\biggl(
W_{{\phi^*_{k}}\,\,\bar c_i\,\,\widehat{\phi_j}\,\,{\phi_{l}}}
+
W_{{\phi^*_{l}}\,\,\bar c_i\,\,\widehat{\phi_j}\,\,{\phi_{k}}}
\biggr )
\nonumber\\&&
+ig^2p_j^\mu \biggl(
W_{{\phi^*_{k}}\,\,\bar c_i\,\,\widehat{A^\mu_j}\,\,{\phi_{l}}}
+
W_{{\phi^*_{l}}\,\,\bar c_i\,\,\widehat{A^\mu _j}\,\,{\phi_{k}}}
\biggr)
\Biggr]
\nonumber\\&&
=\lim_{p_1^2\dots p_4^2=0}
\frac{1}{2}M^2 R^2\sum_{i\not=j}p_i^2 W_{c_i\bar c_i}\Biggl[-
MR
\biggl(\Gamma_{\phi_k\phi_k}W_{b_k\phi_k}
W_{\widehat{\,\,\bar c_{k}}\widehat{\,\, c_i}\,\,\widehat{\phi_j}\,\,\widehat{\,\,\phi_{l}}}
\nonumber\\&&
+
\Gamma_{\phi_l\phi_l}W_{b_l\phi_l}
W_{\widehat{\,\,\bar c_{l}}\widehat{\,\, c_i}\,\,\widehat{\phi_j}\,\,\widehat{\,\,\phi_{k}}}
\biggr )
-ig^2p_j^\mu \biggl(\Gamma_{\phi_k\phi_k}W_{b_k\phi_k}
W_{\widehat{\,\,\bar c_{k}}\widehat{\,\, c_i}\,\,\widehat{A^\mu_j}\,\,\widehat{\,\,\phi_{l}}}
\nonumber\\&&
+
\Gamma_{\phi_l\phi_l}W_{b_1\phi_l}
W_{\widehat{\,\,\bar c_{l}}\widehat{\,\, c_i}\,\,\widehat{A^\mu_j}\,\,\widehat{\,\,\phi_{k}}}
\biggr)
\Biggr]
\label{four.19.7}
\end{eqnarray}
%
 %%%%%%%%%%%%%%%%%%%%%%%%%%%%%%%%%%%%%%%%%%%%%%%%
%%%%%%%%%%%%%%%%%%%%%%%%%%%%%%%%%%%%%%%%%%%%%%%%%
%%%%%%%%%%%%%%%%%%%%%%%%%%%%%%%%%%%%%%%%%%%%%%%%%
%
%
\begin{eqnarray}&&
\nonumber\\&&
=\lim_{p_1^2\dots p_4^2=0}
\frac{1}{2}\bar R\sum_{i\not=j}
\Biggl[-M^4R^4
\biggl(
W_{\widehat{\,\,\bar c_{k}}\widehat{\,\, c_i}\,\,\widehat{\phi_j}\,\,\widehat{\,\,\phi_{l}}}
%\nonumber\\&&
+
W_{\widehat{\,\,\bar c_{l}}\widehat{\,\, c_i}\,\,\widehat{\phi_j}\,\,\widehat{\,\,\phi_{k}}}
\biggr )
\nonumber\\&&
-ig^2
M^3R^3p_j^\mu \biggl(
W_{\widehat{\,\,\bar c_{k}}\widehat{\,\, c_i}\,\,\widehat{A^\mu_j}\,\,\widehat{\,\,\phi_{l}}}
%\nonumber\\&&
+
W_{\widehat{\,\,\bar c_{l}}\widehat{\,\, c_i}\,\,\widehat{A^\mu_j}\,\,\widehat{\,\,\phi_{k}}}
\biggr)
\Biggr].
\label{four.19.7.1}
\end{eqnarray}
%
 %%%%%%%%%%%%%%%%%%%%%%%%%%%%%%%%%%%%%%%%%%%%%%%%
%%%%%%%%%%%%%%%%%%%%%%%%%%%%%%%%%%%%%%%%%%%%%%%%%
%%%%%%%%%%%%%%%%%%%%%%%%%%%%%%%%%%%%%%%%%%%%%%%%%
%
 The final result is then (eqs. (\ref{four.17}),
(\ref{four.18}), (\ref{four.20.1}), (\ref{four.19.1.1})
and (\ref{four.19.7}))
\begin{eqnarray}&&
\frac{1}{M^4}
\lim_{p_1^2\dots p_4^2=0} p_1^\mu p_2^\nu p_3^\sigma p_4^\rho
W_{\widehat{A_\mu}\widehat{A_\nu}\widehat{A_\sigma}\widehat{A_\rho}}
=
R^4\lim_{p_1^2\dots p_4^2=0}
 W_{\widehat{\phi_{1}}\,\,\widehat{\phi_{2}}\,\,\widehat{\phi_{3}}\,\,
\widehat{\phi_{4}}}
\nonumber\\&&
-i\frac{g^2}{2}\lim_{p_1^2\dots p_4^2=0}
R^3\bar R\sum_{i\not=j}\frac{1}{M}
p_j^\mu \biggl(
W_{\widehat{\,\,\bar c_{k}}\widehat{\,\, c_i}\,\,\widehat{A^\mu_j}\,\,\widehat{\,\,\phi_{l}}}
%\nonumber\\&&
+
W_{\widehat{\,\,\bar c_{l}}\widehat{\,\, c_i}\,\,\widehat{A^\mu_j}\,\,\widehat{\,\,\phi_{k}}}
\biggr).
\label{four.19.8}
\end{eqnarray}
{
The second term in the RHS of eq. (\ref{four.19.8}) is zero in the
tree approximation (this is valid in the Landau gauge, while in the 't Hooft
gauge there are tree level diagrams thanks to the direct coupling of the Higgs
boson and the Goldstone boson with the Faddeev-Popov ghosts).
}%%
The dominant term at one loop is the box with
two gauge, one Faddeev-Popov and one Higgs boson propagators shown in Figure 1. Three vertexes
carry a single derivative. Then at high energy the behavior is
$p'_\mu v {\cal O}(\frac{1}{s})$. Thus the total box contribution is
$\sim p^\mu p'_\mu \frac{v}{M}{\cal O}(\frac{1}{s})$,
i.e. of the same order as the first term on the RHS ($\sim 1$).

\begin{figure}
\begin{center}
\includegraphics[width=0.2\textwidth]{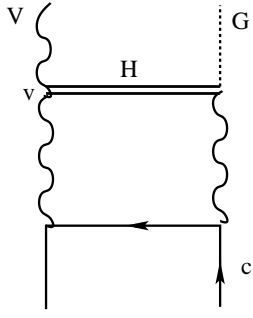}
\end{center}
\caption{One-loop box diagram contributing to the second term in the 
r.h.s. of eq.(58)}
\label{fig.1}
\end{figure}

\subsection{Open Problems}
\begin{itemize}
\item What is the limit theory for $M=0$, if any?
\item In such a limit can we use $v$ as the order parameter?
\item How does the reshuffling of the physical modes occur? In particular,
 does the Goldstone boson become a physical mode?
\item The longitudinal mode $\epsilon_L$ is expected to become
unphysical. How?
\end{itemize}

We should give a second thought to results of Lee, Quigg, Thacker, Weldon,
Chanowitz, Gaillard, Gounaris, K\"ogerler, Neufeld, Denner, Dittmaier, 
Hahn et al. \cite{Denner:1996ug}, \cite{Denner:1997kq}
and look if there is some clue concerning the above listed questions.
Maybe lattice simulations can help in the study of the transition
to $M=0$. These questions might be of great phenomenological significance.
\par\indent
As a conclusion we would dare to say that the above mentioned very
distinguished physicists have extended too much the validity of
their approximations. { In fact, in order to study the very high
energy, they use the set of limiting Feynman rules, that are those
of the massless YM theory, where the longitudinal polarization is an
unphysical mode. }%%

\section{Part Three: Nonlinearly Realized Gauge}

In this part we flash our contribution to the foundation of a
quantum gauge theory, where the group of transformations is realized
nonlinearly.

\subsection{Introduction}
A common structure is present in the nonlinear sigma model (NLSM),
in the massive Yang-Mills (YM) model and in the Higgsless Electroweak model
(EW). For $SU(2)$ one has the action structures: NLSM action (Ref. \cite{0504023}-\cite{0701212})
{
\begin{eqnarray}
S_{\scriptstyle{NLSM}} = \Lambda^{D-4}\frac{M^2}{4}\int d^Dx \,\, Tr\Bigl\{\partial^\mu\Omega^\dagger
\partial_\mu\Omega \Bigr\}
\end{eqnarray}
}
the St\"uckelberg mass for the YM model (Ref. \cite{0705.2339}-\cite{0709.0644})
{
\begin{eqnarray}
S_{YM} \sim \Lambda^{D-4} M^2\int d^Dx \,\, Tr\Biggl\{\Bigl[ A_\mu-i
\Omega
\partial_\mu\Omega^\dagger\Bigr]^2\Biggr\}
\end{eqnarray}
}
and EW (Ref. \cite{0807.3882}-\cite{0903.0281}) mass terms
{
\begin{eqnarray}&& \!\!\!\!\!\!
S_{EW}\sim \Lambda^{D-4}M^2 \int d^Dx \Biggl(
  \,Tr\, \biggl\{\bigl (gA_{\mu}
- \frac{g'}{2} \Omega\tau_3 B_\mu \Omega^\dagger
- i
\Omega
\partial_\mu\Omega^\dagger \bigr)^2\biggr\}
\nonumber\\&&
+
\frac{ \kappa }{2}\Bigl[ Tr\bigl\{gA_{\mu}
- \frac{g'}{2} \Omega\tau_3 B_\mu \Omega^\dagger
- i
\Omega
\partial_\mu\Omega^\dagger \tau_3\bigr\}\Bigr]^2 \Biggr ).
\end{eqnarray}
}
%\normalsize
%
The $2\times 2\in SU(2)$ matrix may be parametrized by the real fields
{
\begin{eqnarray}
\Omega = \phi_0 + i \tau_i \phi_i, \qquad \phi_0=\sqrt{1-\vec\phi^2}.
\end{eqnarray}
}
The constraint is implemented in the path integral measure
{
\begin{eqnarray}
\prod_x{\cal D}^4\phi(x)\theta(\phi_0)\delta(\vec\phi(x)^2+\phi_0^2(x)-1)
=
\prod_x{\cal D}^3\phi(x)\frac{2}{\sqrt{1-\vec\phi^2}}.
\end{eqnarray}
}
The non trivial measure in the path integral is the source of very interesting
facts.

The non polynomial interaction makes the theory nonrenormalizable
{
\begin{eqnarray}&&
S_{\scriptstyle{NLSM}} = \Lambda^{D-4}\frac{M^2}{2}\int d^Dx \,\, \Bigl\{\partial^\mu\phi_0
\partial_\mu\phi_0+ \partial^\mu\vec\phi
\partial_\mu\vec\phi\Bigr\}
\nonumber\\&&
=\Lambda^{D-4}\frac{M^2}{2}\int d^Dx \,\, \Bigl\{ \partial^\mu\vec\phi\partial_\mu\vec\phi+ \frac{1}{\phi_0^2}\,\,\phi_a\partial^\mu\phi_a\,\,
\phi_b\partial_\mu\phi_b\Bigr\}.
\end{eqnarray}
}
Vertexes carry second power of momenta, therefore already at one loop
there is an infinite number of independent divergent amplitudes. Moreover,
 it has been shown in the seventies and in the eighties that some divergences
break chiral invariance (global) at the same order.
\par
\textbf{Strategy:} Abandon Hamiltonian formalism and do perturbation
theory directly on the effective action  functional $\Gamma$.
\subsection{The Local Functional Equation (LFE)}
The measure is invariant under "\emph{local} left multiplication"
transformations $\Omega\to U(\omega(x))\Omega$
{
\begin{eqnarray}&&
\delta \phi_0 = -  \frac{\omega_a(x)}{2}\phi_a
\nonumber\\&&
\delta \phi_a = \frac{ \omega_a(x)}{2}\phi_0 + \frac{
\omega_c(x)}{2}
\epsilon_{abc}\phi_b.
\label{sigma.10}
\end{eqnarray}
}
The following technical work should be done: (i) find the algebra of operators \emph{closed} under
{local} left multiplication transformations by starting from the classical
action, (ii) associate to every composite operator an external classical
source (for subtraction strategy), (iii) write the LFE which follows
from the invariance of the path integral measure.

%%%%%%%%%%%%%%%%%%%%%%%%%%%%%%%%%%%%%%%%%%%%%%%%%%%%%%
%%%%%%%%%%%%%%%%%%%%%%%%%%%%%%%%%%%%%%%%%%%%%%%%%%%%%%
%%%%%%%%%%%%%%%%%%%%%%%%%%%%%%%%%%%%%%%%%%%%%%%%%%%%%%
\subsection*{Step (i) }
This is simple in the NLSM. Introduce the "gauge field"
{
\begin{eqnarray}
F_\mu = \frac{\tau_a}{2}F_{a\mu}\equiv i \Omega\partial_\mu\Omega^\dagger
\label{sigma.10.1}.
\end{eqnarray}
}
Its field strength tensor is zero (it describes a scalar mode) and its transformation
properties are those of a gauge field:
{
\begin{eqnarray}
F_\mu \to U F_\mu U^\dagger + i U \partial_\mu U^\dagger
\label{sigma.10.2}.
\end{eqnarray}
}
The classical action can be written as
{
\begin{eqnarray}
S_{\scriptstyle{NLSM}} = \Lambda^{D-4}\frac{M^2}{4}\int d^Dx \,\, Tr\Bigl\{F_\mu F^\mu \Bigr\}.
\end{eqnarray}
}
Thus the closed set of operators is $\{\vec\phi,\phi_0,\vec F_\mu\}$ .
%%%%%%%%%%%%%%%%%%%%%%%%%%%%%%%%%%%%%%%%%%%%%%%%%%%%%%
%%%%%%%%%%%%%%%%%%%%%%%%%%%%%%%%%%%%%%%%%%%%%%%%%%%%%%
%%%%%%%%%%%%%%%%%%%%%%%%%%%%%%%%%%%%%%%%%%%%%%%%%%%%%%

\subsection*{Step (ii)}
The complete effective action at the tree level is then
{
\begin{eqnarray}
\Gamma^{(0)} = \Lambda^{D-4}\int d^Dx \,\,\Biggl( \frac{M^2}{8}
\Bigl\{F_{a\mu}- J_{a\mu} \Bigr\}^2 + K_0\phi_0\Biggr).
\end{eqnarray}
}
The effective action $\Gamma[\vec\phi,\vec{ J_\mu},K_0]$ is obtained via the Legendre
transform of the logarithm of the path integral functional
{
\begin{eqnarray}
Z[\vec K,\vec{ J_\mu},K_0 ]\equiv \int
\prod_x{\frac{2}{\phi_0}}{\cal D}^3\phi(x)
\exp\Biggl[\Gamma^{(0)}
+\int d^Dy\vec K\vec\phi \Biggr].
\end{eqnarray}
}
\subsection*{Step (iii) }
Now we exploit the invariance of the path integral measure under
local left multiplication ($\delta \phi_a = \frac{
\omega_a(x)}{2}\phi_0 + \frac{ \omega_c(x)}{2}
\epsilon_{abc}\phi_b$).
%%%%%%%%%%%%%%%%%%%%%%%%%%%%%%%%%%%%%%
%%%%%%%%%%%%%%%%%%%%%%%%%%%%%%%%%%%%%%
%%%%%%%%%%%%%%%%%%%%%%%%%%%%%%%%%%%%%%
We expand $\vec \omega(x)$ for small parameter values and obtain the LFE
($\langle\cdots\rangle$ indicates the mean over the weighted paths )
{
\begin{eqnarray}&&
\int d^Dx
\left\langle
\Big(M_D^2(F-J)_{a\mu}(\epsilon_{abc}\omega_cF_b^\mu+\partial^\mu\omega_a)
\right .
\nonumber\\&& \left .
-
\Lambda^{D-4} K_0\frac{\omega_a }{2}\phi_a
+ \phi_0 K_a\frac{\omega_a }{2}+ \epsilon_{abc}K_a\omega_c\phi_b
\Big)(x)\right\rangle=0,
\label{del.3}
\end{eqnarray}
}
where
{
\begin{eqnarray}{
M_D^2\equiv \Lambda^{D-4} M^2.}
\end{eqnarray}
} We will use the notation {
\begin{eqnarray}
{\cal D}[X]_{ab}^\mu= \delta_{ab}\partial_\mu- \epsilon_{abc}X_{c\mu}.
\end{eqnarray}
} Thus for the effective action we get the local functional equation (LFE)
{
\begin{eqnarray}&&
\!\!
-\partial^\mu  \frac{ \delta \Gamma}{\delta J_a^\mu}
+\epsilon_{abc}J_c^\mu\frac{\delta\Gamma}{\delta J_b^\mu}\!\!
+\frac{\Lambda^{D-4} }{2}\phi_a K_0 \!\!+ \frac{1 }{2\Lambda^{D-4}} \frac{ \delta \Gamma}{\delta K_0}
\frac{\delta\Gamma}{\delta \phi_a}
+\frac{1 }{2}\epsilon_{abc} \phi_c  \frac{ \delta  \Gamma}{\delta \phi_b}
%\nonumber\\&&
=0.
\label{del.6}
\end{eqnarray}
}

%%%%%%%%%%%%%%%%%%%%%%%%%%%%%%%%%%%%%%%%%%%%%%%%%%%%%%
%%%%%%%%%%%%%%%%%%%%%%%%%%%%%%%%%%%%%%%%%%%%%%%%%%%%%%
%%%%%%%%%%%%%%%%%%%%%%%%%%%%%%%%%%%%%%%%%%%%%%%%%%%%%%
\subsection{Hierarchy }
The Spontaneous Breakdown of Symmetry is imposed by the condition
{
\begin{eqnarray}\left .
\frac{ \delta \Gamma}{\delta K_0}\right|_{\rm field\,\,\& sources=0}=1.
\end{eqnarray}
} Then the LFE naturally induces a strong hierarchy structure among the
1PI irreducible amplitudes:  {all amplitudes involving the
$\vec\phi$ fields (descendant) are known in terms of the amplitudes
involving only the (ancestor) sources $\vec J_\mu$ and $K_0$}. For
instance, if we differentiate the LFE with respect to
$J_{a'}^\nu(y)$, we get {
\begin{eqnarray}
\!\!\!\!\!\!
\frac{M_D^2}{2}\partial^\mu  \frac{ \delta^2 \Gamma}{\delta J_a^\mu(x)\delta J_{a'}^\nu(y)}
 +  }{\Large
\frac{\delta^2\Gamma}{\delta \phi_a(x)\delta J_{a'}^\nu(y)}}
{
+2\delta_{aa'}\partial_{x^\nu} \delta(x-y)=0.
\label{del.6.2}
\end{eqnarray}
}

%%%%%%%%%%%%%%%%%%%%%%%%%%%%%%%%%%%%%%%%%%%%%%%%%%%%%%
%%%%%%%%%%%%%%%%%%%%%%%%%%%%%%%%%%%%%%%%%%%%%%%%%%%%%%
%%%%%%%%%%%%%%%%%%%%%%%%%%%%%%%%%%%%%%%%%%%%%%%%%%%%%%
\subsection{Weak Power Counting (WPC) }
How many ancestor divergent amplitudes are there ?
The degree of divergence of a graph $G$
for an ancestor amplitude is ($n_L$ is the number of loops)
{
\begin{eqnarray}&&
\delta(G)= D~ n_L -2I +\sum_{j,k}j~V_{jk}+ N_{F}
\nonumber\\&&
n_L=I-\sum_{j,k} V_{jk}-N_F- N_{K_0}+1
\label{wpc.2}
\end{eqnarray}
}
where $I$ is the number of  propagators,
$N_F$   the number of external $F_\mu$ sources and
 $N_{K_0}$ those of ${K_0}$;
$V_{jk}$ denotes the number of vertexes with $k$ $\phi$-lines
and $j$ derivatives.
The superficial degree of divergence $\delta( G)$
for a graph can be bounded by using standard arguments.

%%%%%%%%%%%%%%%%%%%%%%%%%%%%%%%%%%%%%%%%%%%%%%%%%%%%%%
%%%%%%%%%%%%%%%%%%%%%%%%%%%%%%%%%%%%%%%%%%%%%%%%%%%%%%
%%%%%%%%%%%%%%%%%%%%%%%%%%%%%%%%%%%%%%%%%%%%%%%%%%%%%%
By removing $I$ from these  two equations one gets
{
\begin{eqnarray}
\delta(G)= D~ n_L-2n_L  -\sum_{j,k}(2-j)~V_{jk}- N_{F}
-2  N_{K_0}+2.
\label{wpc.3}
\end{eqnarray}
}
The classical action  has vertexes with $j\leq 2$,
therefore, it can be stated that
{
\begin{eqnarray}
\delta(G)\leq  n_L(D-2)+2 - N_{F}-2 N_{K_0}.
\label{wpc.4}
\end{eqnarray}
}
For instance, at $n_L=1$ the only ancestor divergent (independent) amplitudes are
$(J-J)$, $(J-J-J)$, $(J-J-J-J)$, $(K_0-J-J)$ and $(K_0-K_0)$.
The one-loop divergences of graphs where
the descendant field appears ($\vec\phi$) are all expressible all in terms of the ancestor
divergences.

%%%%%%%%%%%%%%%%%%%%%%%%%%%%%%%%%%%%%%%%%%%%%%%%%%%%%%
%%%%%%%%%%%%%%%%%%%%%%%%%%%%%%%%%%%%%%%%%%%%%%%%%%%%%%
%%%%%%%%%%%%%%%%%%%%%%%%%%%%%%%%%%%%%%%%%%%%%%%%%%%%%%

\subsection{Perturbative Expansion}
This is an {\sl Ansatz}. Consider the generic dimension $D$.
Start with $\Gamma^{(0)}$, read from it the value of the
vertexes and construct $\Gamma^{(n)}$ for $n>0$. The connected amplitudes $W^{(n)}$
can then be obtained. Few questions are in order:
\begin{enumerate}
\item Does $\Gamma^{(0)}$ obey the LFE? Yes, by construction
\item Does $\Gamma^{(n)},\, n>0$ obey the linearized LFE ?
{
\begin{eqnarray}&&
\Biggl(
-\partial^\mu  \frac{ \delta }{\delta J_a^\mu}
+\epsilon_{abc}J_c^\mu\frac{\delta }{\delta J_b^\mu}\!\!
+ \frac{1 }{2\Lambda^{D-4}}
\frac{\delta\Gamma^{(0)}}{\delta \phi_a}\frac{ \delta }{\delta K_0}
\nonumber\\&&
+ \frac{1 }{2} \phi_0
\frac{\delta}{\delta \phi_a}
+\frac{1 }{2}\epsilon_{abc} \phi_c  \frac{ \delta  }{\delta \phi_b}
\Biggr)
\Gamma^{(n)}
+\sum_{j=1}^{n-1}\frac{1 }{2\Lambda^{D-4}}
\frac{\delta\Gamma^{(j)}}{\delta \phi_a}\frac{ \delta \Gamma^{(n-j)}}{\delta K_0}
=0.
\end{eqnarray}
}
\item Assume that a {\sl symmetric} subtraction procedure is given for the
divergences in the limit $D=4$. How does the breaking of the above equation occur?
\end{enumerate}

The answers to these questions are given in a compact form by the Quantum Action Principle
{
\begin{eqnarray}&&
\Big(-\partial^\mu \frac{\delta}{\delta J_a^\mu}
+\epsilon_{abc}J_c^\mu\frac{\delta}{\delta J_b^\mu}
\nonumber\\&&
-
\frac{\Lambda^{D-4}}{2} K_0\frac{\delta}{\delta K_a}
+ \frac{1}{2\Lambda^{D-4}} K_a\frac{\delta}{\delta K_0}+ \epsilon_{acb}K_c
\frac{\delta}{\delta K_b}
\Big)Z
\nonumber\\&&
= i
\int
\prod_x{\frac{2}{\phi_0}}{\cal D}^3\phi(x)\Biggl[
-\partial^\mu  \frac{ \delta \hat\Gamma}{\delta J_a^\mu}
+\epsilon_{abc}J_c^\mu\frac{\delta\hat\Gamma}{\delta J_b^\mu}\!\!
\nonumber\\&&
+\frac{\Lambda^{D-4} }{2}\phi_a K_0 \!\!+ \frac{1 }{2\Lambda^{D-4}} \frac{ \delta \hat\Gamma}{\delta K_0}
\frac{\delta\hat\Gamma}{\delta \phi_a}
+\frac{1 }{2}\epsilon_{abc} \phi_c  \frac{ \delta  \hat\Gamma}{\delta \phi_b}
\Biggr]
\exp i\Biggl[\hat\Gamma
+\int d^Dy\vec K\vec\phi \Biggr],
\end{eqnarray}
}
where $\hat\Gamma $ contains the counterterms $\hat\Gamma^{(j)}$,

{
\begin{eqnarray}
\hat\Gamma = \Gamma^{(0)} + \sum_{j=1}^\infty
\hat\Gamma^{(j)}.
\end{eqnarray}
}

\subsection{Subtraction Strategy}

Thus if the counterterms at order $n$ are missing, the linearized LFE
is broken by the term
{
\begin{eqnarray}&&
\Biggl(
-\partial^\mu  \frac{ \delta }{\delta J_a^\mu}
+\epsilon_{abc}J_c^\mu\frac{\delta }{\delta J_b^\mu}\!\!
+ \frac{1 }{2\Lambda^{D-4}}
\frac{\delta\Gamma^{(0)}}{\delta \phi_a}\frac{ \delta }{\delta K_0}
\nonumber\\&&
+ \frac{1 }{2} \phi_0
\frac{\delta}{\delta \phi_a}
+\frac{1 }{2}\epsilon_{abc} \phi_c  \frac{ \delta  }{\delta \phi_b}
\Biggr)
\Gamma^{(n)}
=-\frac{1 }{2\Lambda^{D-4}} \sum_{j=1}^{n-1}
\frac{ \delta \hat\Gamma^{(j)}}{\delta K_0}
\frac{\delta\hat\Gamma^{(n-j)}}{\delta \phi_a}.
\end{eqnarray}
}
Notice that $
{
\frac{1 }{\Lambda^{D-4}}
\frac{\delta\Gamma^{(0)}}{\delta \phi_a}
} $ is independent from $\Lambda^{D-4}$. Thus we use the Laurent
expansion on {
\begin{eqnarray}
\Lambda^{-D+4} \Gamma^{(n)}
\end{eqnarray}
}
to define the finite part and the counterterm ${\Lambda^{-D+4}\hat\Gamma^{(n)}=}$
${-\Lambda^{-D+4} \Gamma^{(n)}\Biggr|_{\rm poles}}$.

The LFE is a power organizer of the divergences that WPC
has classified. The full control can be obtained by finding
the relevant local solutions of the linearized LFE
{
\begin{eqnarray}&&
\Biggl(
-\partial^\mu  \frac{ \delta }{\delta J_a^\mu}
+\epsilon_{abc}J_c^\mu\frac{\delta }{\delta J_b^\mu}\!\!
+ \frac{1 }{2\Lambda^{D-4}}
\frac{\delta\Gamma^{(0)}}{\delta \phi_a}\frac{ \delta }{\delta K_0}
\nonumber\\&&
+ \frac{1 }{2} \phi_0
\frac{\delta}{\delta \phi_a}
+\frac{1 }{2}\epsilon_{abc} \phi_c  \frac{ \delta  }{\delta \phi_b}
\Biggr)
\Gamma^{(n)}[\vec\phi,\vec J_\mu,K_0]
=0.
\end{eqnarray}
} This can easily be achieved by using the technique of {\sl
bleaching}. We shortly describe this procedure. The above equation
naturally suggests the following infinitesimal transformations:
%$\delta_0$.
%
%The transformations
{
\begin{eqnarray}&&
\delta_0 J_b^\mu = (\partial^\mu\delta_{ab}  +\epsilon_{abc}J_c^\mu)\omega_a={\cal D}[J]_{ba}^\mu\omega_a
\nonumber\\&&
\delta_0 F_a^\mu = {\cal D}[F]_{ab}^\mu\omega_b
\nonumber\\&&
\delta_0 K_0 = -
\frac{\omega_a }{\Lambda^{D-4}}
\frac{\delta\Gamma^{(0)}}{\delta \phi_a}
\nonumber\\&&
\delta_0 (-\frac{\delta\Gamma^{(0)}}{\delta \phi_a})=
\Lambda^{D-4}\frac{1}{2} \omega_{a} K_0
+\frac{1}{2} \epsilon_{abc}  \omega_c (-\frac{\delta\Gamma^{(0)}}{\delta \phi_b}) \,,
\end{eqnarray}
}
 which lead to the bleaching
{
\begin{eqnarray}&&
{\mathfrak J}_{\mu} \equiv \Omega^\dagger (J_{\mu}-F_{\mu})\Omega
\nonumber\\&&
{\mathfrak K}_0 \equiv \frac{K_0}{\phi_0} -\frac{M^2}{4}(J_b^\mu-F_b^\mu)\frac{\partial F_{b\mu}}{\partial \phi_a}\phi_a
\end{eqnarray}
}
Here are few facts about bleaching. i) The relations are invertible, ii) In
the case of ${\mathfrak J}_{a\mu}$, bleaching  is a kind of gauge
transformation where the parameters are the $\vec\phi$ fields:
{
\begin{eqnarray}&&
{\mathfrak J}_{\mu} =\Omega^\dagger J_{\mu}\Omega + i \Omega\partial_\mu\Omega
\nonumber\\&&
\partial_\mu{\mathfrak J}_{\nu} =\Omega^\dagger \left(\partial_\mu+
\Omega\partial_\mu\Omega^\dagger
\right)(J_{\nu}-F_{\nu})\Omega=\Omega^\dagger{\cal D}_\mu[F](J_{\nu}-F_{\nu})\Omega
\end{eqnarray}
} iii) the invariants can be constructed by using ${\mathfrak
J}_{\mu}$ and ${\mathfrak K}_0 $ and their space-time derivatives.

Ancestor amplitudes do not depend explicitly on $\vec\phi$. We
consider only those relevant for the one-loop divergences.
\par\noindent
We give here a list of the relevant one-loop invariants necessary
for the parameterization of the one-loop divergences of the NLSM:
%
%The one-loop invariants are:
%
{
\begin{eqnarray}
&& {\cal I}_1 = \int d^Dx \, \Big [ D_\mu ( F -J )_\nu \Big ]_a \Big [ D^\mu ( F -J )^\nu \Big ]_a  \, ,
\nonumber \\
&& {\cal I}_2 = \int d^Dx \, \Big [ D_\mu ( F -J )^\mu \Big ]_a \Big [ D_\nu ( F -J )^\nu \Big ]_a  \, ,
\nonumber \\
&& {\cal I}_3 = \int d^Dx \, \epsilon_{abc} \Big [ D_\mu ( F -J )_\nu \Big ]_a \Big ( F^\mu_b -J^\mu_b \Big ) \Big ( F^\nu_c -J^\nu_c \Big ) \, ,  \nonumber \\
&& {\cal I}_4 = \int d^Dx \, \Big ( \frac{K_0}{\phi_0} +\frac{M^2}{4}[F_b^\mu-J_b^\mu]\frac{\partial F_{b\mu}}{\partial \phi_a}\phi_a \Big )^2 \, , \nonumber \\
&& {\cal I}_5 = \int d^Dx \, \Big ( \frac{ K_0}{\phi_0}  +\frac{M^2}{4}[F_b^\mu-J_b^\mu]\frac{\partial F_{b\mu}}{\partial \phi_a}\phi_a  \Big ) \Big ( F^\mu_c -J^\mu_c \Big )^2 \, ,
\nonumber \\
&& {\cal I}_6 = \int d^Dx \, \Big ( F^\mu_a -J^\mu_a\Big  )^2
 \Big ( F^\nu_b -J^\nu_b \Big )^2 \, , \nonumber \\
&& {\cal I}_7 = \int d^Dx \, \Big ( F^\mu_a -J^\mu_a\Big  )
   \Big ( F^\nu_a -J^\nu_a\Big  )
%\nonumber \\&& ~~~~~~~~~~~~~~~~
   \Big ( F_{b\mu} -J_{b\mu} \Big  )
   \Big ( F_{b\nu} -J_{b\nu} \Big  ) \, ,
\end{eqnarray}
}
where $D_\mu$ denotes the covariant derivative w.r.t $F_{a\mu}$:
\begin{eqnarray}
D_{ab\mu} = \delta_{ab}\partial_\mu  - \epsilon_{abc} F_{c\mu } \, .
\label{appE:5}
\end{eqnarray}

The counterterms are evaluated by extracting the pole parts from the
relevant amplitudes given by the effective action functional normalized
by
{
%
%\begin{eqnarray}
$
\Lambda^{-D+4}\Gamma.
$
%\end{eqnarray}
%
}
It is very important to care about the relation

{
\begin{eqnarray}
\!\!\!\!
2({\cal I}_1 -{\cal I}_2 )-4{\cal I}_3+({\cal I}_6-{\cal I}_7 )= \int d^D x {\cal G}_{a\mu\nu}[{\mathfrak J}]{\cal G}_a^{\mu\nu}[{\mathfrak J}]
%\nonumber\\&&
= \int d^D x {\cal G}_{a\mu\nu}[J]{\cal G}_a^{\mu\nu}[J]=\sim 0.
\end{eqnarray}
}

The last integral is sterile: no descendant terms are generated.
Now the calculation gives
{
\begin{eqnarray}
%&&
 \G^{(1)}
= \frac{1}{D-4} \frac{\Lambda^{D-4}}{(4\pi)^2}\Big [
- \frac{1}{12}   \Big (
{\cal I}_1 - {\cal I}_2 -  {\cal I}_3 \Big )
%\nonumber \\& &
+  \frac{1}{48}
\Big ( {\cal I}_6 + 2 {\cal I}_7 \Big )
%\nonumber \\& &
+
 \frac{3}{2} \frac{1}{M^4 } {\cal I}_4
+  \frac{1}{2} \frac{1}{M^2} {\cal I}_5 \Big ] \, .
\end{eqnarray}
}

\subsection{The massive Yang-Mills theory}
$\Omega$ describes the Goldstone bosons, that are here unphysical modes.
Then it is important to ensure that the Slavnov-Taylor Identity (STI) is valid
in order to preserve unitarity.
The LFE must be compatible with the STI. A suitable gauge-fixing term will help
to achieve this result. The Landau gauge is the simplest, since the tadpole
contributions can be neglected in most cases. The transformations to be considered
are the local left $SU(2)_L$ and the global right $SU(2)_R$ on $\Omega$, the gauge fields are
 $A_\mu$ and the Faddeev-Popov fields are $c, \bar c$. Few external sources are needed in order
to describe the complete (under the $SU(2)_L \otimes SU(2)_R$) set of composite operators.

The action in the presence of the Landau gauge-fixing terms looks as
follows:
\begin{eqnarray}
\G^{(0)} & = &
 S_{\scriptstyle{YM}} +
\frac{\Lambda^{D-4}}{g^2}
\int d^D x \, \Big ( B_a (D^\mu[V](A_\mu - V_\mu))_a
- \bar c_a (D^\mu[V] D_\mu[A] c)_a \Big )
\nonumber \\& &
+ \int d^Dx \, \Big ( A^*_{a\mu} sA^\mu_a +
\phi_0^* s \phi_0 +
\phi_a^* s \phi_a + c_a^* s c_a
+ K_0 \phi_0 \Big ) \, .
\label{brst.4.2}
\end{eqnarray}
\begin{eqnarray}
S_{\scriptstyle{YM}}
   = \frac{\Lambda^{(D-4)}}{g^2} \int d^Dx \, \Big ( - \frac{1}{4}
                              G_{a\mu\nu}[A] G^{\mu\nu}_a[A] +
                    \frac{M^2}{2} (A_{a\mu} - F_{a\mu})^2 \Big ) \, .
\label{stck.1}
\end{eqnarray}
\begin{eqnarray}
\Omega = \frac{1}{v} ( \phi_0 + i \tau_a \phi_a ),\qquad \phi_0^2 + \phi_a^2 = v^2  \,
\label{s2.2}
\end{eqnarray}
where $v$ is a parameter with dimension one. We stress that
$v$ is {\emph{not a parameter of the model}}, because it can can be
removed by a rescaling of the fields $\vec\phi$ and $\phi_0$.

\subsection*{Slavnov-Taylor Identity}
%
%The Slavnov-Taylor Identity necessary for the validity of physical unitarity:
 The $S$-matrix  satisfies the following equation at the {\emph{perturbative level}}:
$$
\langle\alpha  |\beta\rangle =
\sum_{n \in \{{\rm physical\,\, states}\}}\langle\alpha  | S |n\rangle \langle n  | S^\dagger |\beta\rangle
$$
if both $\alpha$ and $\beta$ are physical states. This in general is valid if the Slavnov-Taylor identity  is valid.
{
\begin{eqnarray}
%&& \!\!\!\!\!\!\!\!\!\!
{\cal S}(\G) = \int d^Dx \, \Big (
\frac{\delta \G}{\delta A^*_{a\mu}} \frac{\delta \G}{\delta A_a^\mu}
+
\frac{\delta \G}{\delta \phi_a^*} \frac{\delta \G}{\delta \phi_a}
+
\frac{\delta \G}{\delta c_a^*}\frac{\delta \G}{\delta c_a}
+ B_a \frac{\delta \G}{\delta \bar c_a} 
%\nonumber \\
%&& ~~~~~~~~~~~~~~~~
      - K_0 \frac{\delta \G}{\delta \phi_0^*}
\Big ) = 0 \, .
\label{brst.13p}
\end{eqnarray}
}
The LFE for the massive YM model can be cast in the form:
%
%The Local Functional Equation for the case of massive Yang-Mills
%theory has the following form:
%
%
{
\begin{eqnarray}
&&{\cal W}(\G) \equiv \int d^Dx \alpha_a^L(x)\Biggl(
-\partial_\mu \frac{\delta \G}{\delta V_{a \mu}}
+ \epsilon_{abc} V_{c\mu} \frac{\delta \G}{\delta V_{b\mu}}
-\partial_\mu \frac{\delta \G}{\delta A_{a \mu}}
\nonumber \\&&
+ \epsilon_{abc} A_{c\mu} \frac{\delta \G}{\delta A_{b\mu}}
+ \epsilon_{abc} B_c \frac{\delta \G}{\delta B_b}
+ \frac{1}{2} K_0\phi_a
{\Large
+ \frac{1}{2} \frac{\delta \G}{\delta K_0}
\frac{\delta \G}{\delta \phi_a}
}
\nonumber \\
&&
+
\frac{1}{2} \epsilon_{abc} \phi_c \frac{\delta \G}{\delta \phi_b}
      + \epsilon_{abc} \bar c_c \frac{\delta \G}{\delta \bar c_b}
      + \epsilon_{abc} c_c \frac{\delta \G}{\delta  c_b}
 \nonumber \\
&&
      + \epsilon_{abc} A^*_{c\mu} \frac{\delta \G}{\delta A^*_{b\mu}}
      + \epsilon_{abc} c^*_c \frac{\delta \G}{\delta  c^*_b}
 + \frac{1}{2} \phi_0^* \frac{\delta \G}{\delta \phi^*_a}
\nonumber \\
&& +
\frac{1}{2} \epsilon_{abc} \phi^*_c \frac{\delta \G}{\delta \phi^*_b}
- \frac{1}{2} \phi_a^* \frac{\delta \G}{\delta \phi_0^*}
\Biggr)
= 0 \, .
\label{bkgwi}
\end{eqnarray}
}
$\G$ also obeys the Landau gauge equation
\begin{eqnarray}
\frac{\delta \G}{\delta B_a} = \frac{\Lambda^{D-4}}{g^2}
 D^\mu[V](A_\mu - V_\mu)_a
\label{b.eq}
\end{eqnarray}

\subsection*{Linearized Equations and Induced Transformations}
The structure of both STI and LFE is standard. Thus we can
\begin{enumerate}
\item Establish the full hierarchy (only the Goldstone bosons are descendant fields)
\item Confirm the validity of the WPC
\item Introduce the linearized STI and LFE
\item Extract from the linearized STI and LFE the generators of the
transformations on the effective action $\Gamma$
\item Check that the generators stemming from STI commute with those from LFE
\end{enumerate}

\subsection*{Subtraction procedure}

With these tools we can construct the most general classical action
compatible with the WPC and the invariance under the BRST
transformations and the LFE induced symmetry. Surprisingly enough,
 the resulting action is the standard YM field theory with a
St\"uckelberg mass term.
\par
The subtraction procedure of the divergences is then the same as in the NLSM:
subtraction of the pure pole parts in the Laurent expansion around $D=4$
of the normalized amplitudes $\Lambda^{-D+4}\Gamma$.
This subtraction procedure has been implemented in the one-loop
calculation of the gauge field two-point functions \cite{0709.0644}, \cite{0903.0281}.
Moreover, it has been tested for a solvable model \cite{0907.0426}.

\subsection*{Consistency of the Subtraction Procedure }
The two-loop self-energy amplitude
has been considered from the point of view of the consistency.
It has been argued that the subtraction scheme is consistent:
i) the counterterms are local ii) physical unitarity is satisfied
iii) the STI and LFE induced symmetry on $\Gamma$ is preserved.

In Ref. \cite{0709.0644} we proved the following results:
\par\noindent
1) explicit calculation of the gauge field two-point function.
\par\noindent
2) Check that the counterterms are local at the two-loop level.
\par\noindent
3) Validity of unitarity.
\par\noindent
4) All divergences (infinite) at the one-loop level are subtracted by a finite number
of counterterms.

\subsection*{Outlook and (some) open questions}
Several issues should be addressed:
\begin{itemize}
\item Phenomenological applications
\item Running constant (dependence on $\Lambda$)
\item How to proceed with a generic regularization tool?
\item Well-defined strategy of minimal subtraction with
      anticommuting $\gamma_5$.
\item Extension to Grand Unified groups
\end{itemize}

\section{Conclusions}

Our approach to theories with nonlinearly realized gauge group
is based on the Local Functional Equation, which applies to the generating
functionals. The features of this method are quite novel in field theory
and can be briefly summarized as follows:
\begin{itemize}
\item Hierarchy: all the amplitudes involving the parameter fields (the pion field in the
nonlinear 
sigma model, the Goldstone bosons in the nonabelian gauge theories) can be derived from
well-defined ancestor field amplitudes given in terms of gauge-  and order-parameter-fields.
This property allows one to fix at every order an infinite number of divergent amplitudes
in terms of a finite number of divergences involving only the ancestor fields.
\item Weak Power Counting: for the ancestor amplitudes a criterion is needed in order
to make hierarchy effective. The subtraction procedure that we are implementing
is compatible with the WPC, i.e. if the starting action is constructed by using
the WPC, then the counterterms do not alter this property.
\item Existence of a consistent subtraction procedure (symmetric and local): it can be proven
that minimal dimensional subtraction on properly normalized amplitudes maintains the validity
of the LFE.
\item Necessity of a finite number of physical parameters. It is essential that the number of
free parameters is finite and independent from the order in the loop expansion. Otherwise the
subtraction strategy would not be consistent, since every parameter should be present in the
tree-level action. 
\end{itemize}
For massive Yang-Mills theory, using Slavnov-Taylor identities and
the Landau gauge equation we proved
\begin{itemize}
\item The physical unitarity of the theory. This property is of paramount importance since
our approach, as in the usual linear case, has unphysical modes (Goldstone bosons, spin-zero
vector field polarization, Faddev-Popov ghosts). The proof proceeds in the standard way
by showing that the unphysical modes cancel in the unitarity equation for the S-matrix
involving only physical states.
\item The consistency of the Local Functional Equation with all other equations, such as
the Slavnov-Taylor identities, the gauge-fixing equation and the anti-ghost equation.
All the equations are not spoiled in the
proposed subtraction procedure.

\item We finally mention that the massive YM theory can also be
formulated in the 't Hooft-Feynman gauge. However, in this gauge
one has to deal with many tadpole diagrams that are absent
in the Landau gauge.
\end{itemize}

\section*{Acknowledgments}
{
One of us (R.F) is pleased to thank the Center for Theoretical Physics at MIT,
Massachusetts, where he had the possibility to work on this research. This work is supported in part by funds provided by the U.S. Department
of Energy (D.O.E.) under cooperative research agreement \#DE FG02-05ER41360.}

\end{document}